\documentclass[10pt,a4paper]{article}
\usepackage[utf8]{inputenc}
\usepackage[T1]{fontenc}
\usepackage{amsmath}
\usepackage{amsfonts}
\usepackage{amssymb}
\usepackage{graphicx}
\usepackage{fullpage}

\begin{document}
\title{Quantum Mechanics on a background modulo observation}	
\author{Jose A. Pereira Frugone \\ japfrugone@yahoo.com}

\maketitle

\textbf{Abstract} \\
In this work we will answer the following question: What remains of Quantum Mechanics when we  transform the background space-time into a space modularized by observation or measurement regions ? This new moduli space is constructed by identifying regions of space-time where quantum phase comparison (observation, measurement) is implied. We call it Observation Modular space (OM-space). In addition we replace in QM statements the Plank constant (h) by the quantity $\zeta_0 4 \pi^2$ (where $\zeta_0$ is the Plank Length) or otherwise, replacing $P_0$ (the Planck Momentum) by $4 \pi^2$. This maps Quantum Mechanics into a very rich dual Number Theory which we call Observation Modular Quantum Mechanics (OM-QM). We find  the  OM-dual to the Dirac Equation, the quantum Wave Function and a free particle's mass. The  OM-QM counterparts of the Energy turns out to be a simple function of the zeroes of the Riemann zeta function. We also find the OM-QM correspondents to the electron spin, the electron charge, the Electric Field and the Fine Structure Constant. We also find the OM-QM correspondents of the Heisenberg uncertainty relation and Einstein's General Relativity Field equation emerging as certain limits of a unique OM-QM equation. We also get the OM-QM correspondents of the Gravitational Constant and the Cosmological Constant. We find the analog of holography in the OM-QM side and we get an interpretation of spin as a high dimensional curvature. An interpretation of the OM-QM correspondence is proposed as giving the part of QM information which is not measurement or observation dependent. Some potential future applications of this correspondence are discussed.

\section{Introduction}
In this work we will ask the following question: What remains of Quantum Mechanics when we identify (modularize) regions of the space-time background where observation or measurement is implied ? To be precise we are not talking here about the Observer itself as this is a more complex topic which in general is not well understood and difficult to model inside Quantum Mechanics (QM). Neither we are talking here about the Measurement or Observation process itself which is also a very hard problem in QM.  One would think (naively) that according to the Copenhagen interpretation of QM the answer to this question would be : Nothing at all remains except from Mathematics disconnected from QM physical content. According to this interpretation (the most standard and accepted one) QM is only talking about the information one can get in measurements involving an observed system. Thus, if we strip QM of observation (represented by a concrete physical mechanism like quantum wave function phase comparison between different regions of space-time) there should be no physically meaningful information content left in the residual Theory. That is, naively we could think that if we reduce QM over a moduli space defined modulo those regions where that observation or quantum phase comparison is happening, the resulting Theory should have no structure left other than probably pure Mathematics. Here we will prove that, contrary to that naive conclusion, we can construct a very specific such modulli space for which the reduction of QM on it is highly structured and non-trivial. We will call this moduli space Observation Modular space (OM-space ). Moreover, this resulting Theory while being purely Mathematical heritages (by design) a lot of the interconnections between Physical observables and makes evident surprising new interconnections and insights coming from the mathematical structures in it. Although one could define such modularization in apparently infinite ways, we use a very specific parametrization of the target theory which gives very suggestive and meaningful interconnections between Mathematical objects matching the ones in the Physical side. We call the residual Theory resulting after this parametrization and modularization Observation Modular Quantum Mechanics (OM-QM ). The intuition which takes us to this special parametrization is the following: \\ \\
The number $4 \pi^2$ (a transcendental number) and the Physical quantity h (the Quantum of Action or Planck's constant) play a  similar role in Number Theory and in Physics respectively. $ 4 \pi^2 $ is connected to the distribution of ensembles of prime numbers. It is the result of the multiplication of all prime numbers "alla Euler" \cite{munoz2003product}, that is, via Analytical Continuation of that infinite product. It's in that sense that we should understand the statement that $4 \pi^2$ "knows" about the distribution of ensembles of prime numbers. \\  By the other hand the Planck constant encodes the information about the distribution of ensembles of quantum states in an arbitrary physical system \cite{dirac1926theory}. In both cases we are talking about statistical distributions of certain irreducible elements of the theory. \\  In the case of h that relationship with the distribution of quantum states is more "by design". From a mathematical point of view the Dirac axiom of QM is a statement about the relative distribution of eignenvalues of certain non-commuting, Hermitean operators (the physical observables) acting on a Hilbert space. h is the only parameter in such distribution \cite{cohen2019quantum}. \\
We argue here that there is more than an analogy between the roles of those two constants. We conjecture that every statement in Quantum Mechanics can be mapped into a corresponding Number Theoretical statement  which results from replacing h with the quantity  $\zeta_0 4 \pi^2$ (where $\zeta_0$ is the Planck Length) while at the same time modularizing the Space-Time background by identifying regions where quantum phase comparison is implied. \\
Another way to say the same is that there is a correspondence between a Quantum Mechanical statements and a certain Number Theory statement which results from replacing the Planck's momentum, defined as : \\ \\
\begin{equation}\label{Eq1}
P_0 = \frac{h}{\zeta_0} 
\end{equation} \\ \\ 

 with the number $4 \pi^2$ and transforming Space-Time in the aforementioned way. We will call this reparametrization the $P_0$-$4 \pi^2$ correspondence. We will call the full process of transforming a QM statement into its OM-QM dual the OM-correspondence. This process amounts to reducing Quantum Mechanics over a moduli-space background where the modularity is given by identifying regions of space-time where observation or measurement is implicit. We will define and explore this OM-space in the rest of this work. 
 
The effects of such OM-correspondence on QM seems to be fine tuned to give us a dual highly non-trivial Number Theory surprisingly full of physical and mathematical interconnections. We will analyze its internal consistency as well as potential connections with several similar ideas which have appeared in the past in the literature. Indeed, monduli spaces in relation to quantum based theories are not a new topic. The subject of Topological Quantum Field Theories (QFT) develops similar ideas but in a Field Theory context. In QFT the interest is on finding the effects on a Quantum Field Theory when one replaces normal space-time with a moduli-space constructed via some kind of modularity related to the Gauge Symmetries in the Fields \cite{witten1988topological} \cite{witten1988space}. In QFT the starting point is an action which is independent on the space-time metric. Those actions are dependent only in certain functions of the field potential which give raise to Feynman amplitudes which in turn are topolocical invariants (Knot invariants sometimes) of the moduli space . \\ A reference which is even more relevant for our work is Gao's work on AdS/CFT correspondence over a quotient space \cite{gao1999ads}. In it a moduli (quotient) space version of the bulk AdS space is introduced. The modularization in this case is via some large, general quotient Group. Gao proceeds then to find the effects of this modularization in the boundary quantum CFT Theory using the AdS/CFT correspondence. This is very similar to what we will attempt in our work. What is new in our approach is that we are in general interested in the effects of that modularization in Quantum Mechanics iteslf in a more realistic setup. We are not going to base on a toy model bulk space-time background as AdS but instead base in the usual 4-Dimensional space time and in full Quantum Mechanics. Neither we will use a large group for producing a quotient space but instead we will use observation regions of space as our quotient criteria. Also it is new the $P_0$-$4 \pi^2$ substitution we use in our OM-correspondence. The main conclusion we will be lead to is that the OM-QM construction reduces the degrees of freedom available in Quantum Mechanical information in a very special way. It leaves us with a Physics independent core of that information which itself contains a rich structure. \\ We can say that that residual Number Theory core we are left with is the Observation independent content of Quantum Mechanical information. We will see that topics which are difficult or obscure in the usual QM framework become natural and easy to deal with in the OM-QM side. Topics in QM do get mapped to very concrete and clear Mathematical  statements which often throw some light on their nature in the QM side.\\ In Mathematics we often see similar simplification or insight when we pass some problems to a modular framework. OM-QM can be seen as the counterpart of that Mathematical modular simplification or insight in QM.\\ We will first apply the OM-correspondence to the Quantum Mechanics of a free electron. That is, we are going to find the OM-dual of the Dirac Equation. This will gives as a by-result that the OM-QM  counterparts of the particle mass and energy are simple functions of the zeroes of the Riemann Zeta function. Then we will use this correspondence for finding a fist principles derivation of the value of the Fine Structure Constant (it precise numeric value!) \\ \\ In this work we will adopt a Physics approach for developing the topics at hand. We will not delve too deep in the Mathematics aspects although we will be dealing with deep Mathematical concepts. For example when talking about moduli-spaces one could adopt a very strict Mathematical approach as that is in reality a very complex Mathematical subject \cite{atiyah1988topological}. We will use here the wording moduli-spaces as a placeholder for an eventual future formalization of the Physical ideas we will propose. We will go as deep as we need in the Mathematics but not so deep that we lose focus on our main goal, finding a precise definition of the OM-correspondence and proving that it produces an interesting and very non trivial effect in QM. The Theory we will develop below seems to be a small common playground where Physics and Mathematics can play together for the sake of finding new or known interconnections. We propose that OM-QM is in fact a toy model of a Quantum Gravity Theory. In the final section of this work we will briefly talk about potential future applications o this correspondence.

\section{The OM correspondence}
The correspondence we are postulating in this paper can be summarized in the following procedure: \\ \\
\begin{itemize}

\item[\bf{step 1}] Take a valid Quantum Mechanical statement (an equation, a definition, a postulate). Let's call this the $P_0$ side of the correspondence. The OM corresponding statement can be constructed via the following operations:\\
\item[\bf{step 2}] Replace any instance of the Planck constant h by the quantity $\zeta_0 4 \pi^2$ (where $\zeta_0$ is the Planck Length). Or equivalently, replace $P_0$ (the Planck Momentum) in the QM side by the number $4 \pi^2$ \\
\item[\bf{step 3}] Identify all regions in the background space-time where a quantum phase comparison (that is measurement or observation) is implicit in the quantum statement (up to homotopy classes). Keep the form of the QM equations covariant in the transformation. Then find the effect of such transformation in all elements in the quantum statement. We will make a unique and precise definition of this modularization in next section.
\end{itemize}

In the rest of the paper we will notate with a tilde the OM-QM correspondent of any quantum quantity. Then by definition we have 

\begin{equation}\label{Eq2}
\tilde{P_0} = 4 \pi^2 
\end{equation}   

We will also make use of terms like "OM-Energy" or "OM-mass" for referring to the OM correspondents of certain Physical observables. 

\section{OM-space: Space-Time modulo regions with implied Quantum wave function phase comparison}

Let's analyze the step 3 in the above mentioned procedure. As one of the elements of that step we impose covariance of the Quantum equations under the space-time transformation we are performing. That is, we are imposing that the form of the dual Number Theoretic statement should be the same as the form of the Quantum Mechanical statement. We know which are the space-time transformations under which Quantum Mechanics is covariant. Ignoring (by now) the presence of any gauge fields in the Theory, in the case of non-relativistic QM those are the transformations from the Galileo Group. In the case of relativistic QM those are the Lorentz group transformations multiplied by the internal spin symmetry group. These internal spin symmetry groups are representations of the SU(2) group acting on 4 components spinors. The internal symmetry groups can include Gauge groups if we are talking about the QM of particles interacting with Gauge fields \cite{dirac1932relativistic} \cite{greiner2000relativistic} \cite{wilczek1999quantum}. This covariance enforces the elements of the QM equations to transforms in very specific ways. For example mass must transform as scalar  under Lorentz transformations. Gauge charges (electric, color, lepton flavors) will transform as vectors under the internal gauge groups. The particle states will in turn transform as spinors in the relativistic case and under the internal gauge groups if gauge charges and fields are considered. The space-time transformation we use in Step-3 of our correspondence is obviously none of those transformations under which QM is covariant. Therefore if we want to impose covariance under our space-time transformation the way the QM elements transform must be different than in the way QM transforms under its covariance transformations. We must assume that all elements in the QM equation must transform in general in order to keep covariance. Elements as Mass, Gauge Charges (electric of other), Spin and the state vectors will all change in these transformations in ways we will find in the rest of this work. In fact Step-3 is all about the fact that even when we are imposing that the form of the equation should not change then all its elements should change to keep that covariance. Finding out how those elements change in order to keep covariance is the core of this work. In this work we are indeed talking about a type of covariance which is not Lorentz or Gauge covariance. We argue that this covariance is related to the observation itself. We are imposing that the QM laws should keep their form even when we take out  observation or measurements out from the picture. Or otherwise, what we are defining here  is QM Modulo Observation.\\ \\

Although Step-3 in our OM correspondence is a simple definition from the physical point of view it is not very practical if one wants to apply it in an operational way to real quantum mechanical statements of any kind. If we want to apply this correspondence in practice we need to describe in a more precise and operational way what we mean by modularizing the Phase Space by identifying regions where quantum phase comparison is implicit. We will use a similar reasoning to the one  Maldacena and Susskind employed in their ER=EPR correspondence \cite{maldacena2013cool} \cite{susskind2016copenhagen}. They needed to make more quantitative their principle of identifying regions of space-time connected via entanglement. In their case they studied the correlations between states in entangled Space-Time regions and compared them to the correlations of events in Space-Time regions connected via an Einstein-Rosen bridge. They mapped the "gluing" of those regions into a more operational notion: probability correlation of states in those regions. \\
For the OM correspondence we need a similar more operational form of Step 3. We will not try to find at this point an explicit form of the Phase Space transformation implied in that step (we will do that later in this work). We will define instead our moduli Space-Time in a totally equivalent way from the Quantum Mechanical point of view. It is enough describing how such transformation of the space-time acts on Integrals over paths in Phase Space.\\ \\ In Feynman's  interpretation of QM, the central object of the Theory is the quantum Amplitude given by the Feynman path integral \cite{feynman2010quantum}

\begin{equation}\label{Eq3}
\mathcal{A}_{\alpha \rightarrow \beta} = \int_\mathbf{C} ( \exp{ ( \frac{i*S}{\hbar} ) } ) ds 
\end{equation}

where $ \mathbf{C} $ is the set of all possible paths in Space-Time taking us from an initial point $ \alpha $ to a final point $ \beta $  and S is the Classical Action for the system. All the information about results of measurements is contained in that amplitude. According to the Copenhagen interpretation of QM that is all the information we can expect to know about quantum systems. Therefore, if we find what is the effect of the modularization of Phase Space described in Step 3 on that Amplitude we will have an equivalent and more operational form of that Step 3 in the OM correspondence. \\  \\

\begin{figure}[h]
	\centering
	\includegraphics[width=0.8\linewidth]{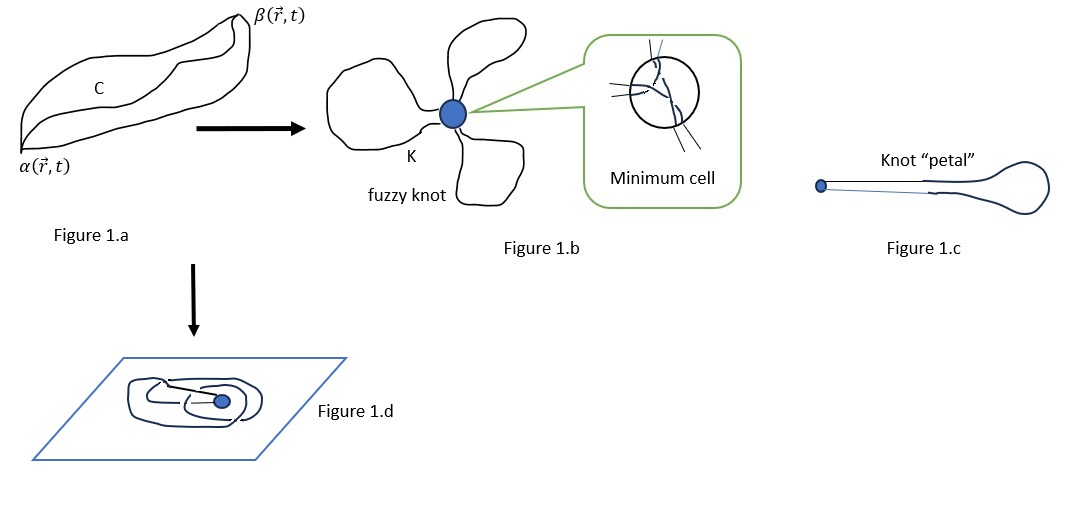}
\end{figure}

Here we make a crucial observation about the end points of the set of paths $ \mathbf{C} $ (the start classical space-time point $ \alpha $ and end point $ \beta $ ). There is a quantum phase comparison assumed in each of them. We  motivate this by saying that the fact we are talking about those two extreme points for all the paths carries the assumption that there is a measurement implicit at both ends. Nevertheless, the value of the Feynman amplitude does not depend on a measurement or observation of one concrete observer over the system. According to the Copenhagen interpretation of QM the only thing this Amplitude is saying is that if one performs the experiment or observation for measuring the probability of transition of the system from a state centered (according to the uncertainty dispersion) in the space-time point $ \alpha $ to a state centered in space-time point $ \beta $ then the value of such probability is given by Eq.\ref{Eq3}. In this Amplitude formula we are always speaking about information from experiments or observations. There is observation implicit in it. The set of paths $ \mathbf{C} $ is in general infinite and other than the experimental constraints there is no limitation in how long or convoluted they can be. In fact QM says we have to integrate over all of the possible paths. The potential observational information in the set of paths comes from two sources: from the common end-points and from some experimental or topological constraints implicit in the set up of the Classical system. For example if we are performing the double slit experiment, there are obvious topological/geometrical constraints implicit in the measurement and in the amplitude. In addition we have the constraints coming from the extreme points of the the paths. In Step 3 of the OM correspondence we explicitly mention that the identification of regions of the space-time is up to homotopy classes. This topological reduction or identification of paths accounts for the observation of  topological constraints in the system. But we still have to take into account the part of the amplitude  implicit in the end points of all the paths. We argue here that this localization of the end points are equivalent to a measurement. In fact we argue that any kind of constraint on the system should be considered as assumed   measurement and therefore a quantum phase comparison. In any measurement or constraint a quantum phase comparison is implicit. Therefore following Step 3 we can identify both end points  $ \alpha $ and  $ \beta $ up to a cell of minimal size in space-time (a pixel given by Heisenberg's uncertainty relation). Going forward, we will consider that minimal size pixel to be a Planck Length-Time radius ball in 4 dimensions. The net result of this reduction of number of paths due to topological equivalence and to the identification of the end points as described above is that the set of paths $ \mathbf{C} $ gets mapped to a finite set of closed loops united in a fuzzy way on that minimal size pixel of space-time. \\ \\

In Figure 1 we describe this transformation in a pictorial way in 2-D for the sake of visualization. First thing one can notice in Figure 1.c is the way the transformed end parts of the paths "approach" the minimal size cell. They enter the region close to the minimal cell in an almost parallel way. Not totally  parallel but the closer they get to the minimum cell the more parallel they become. The reason for this is the following: If this transformation is a proper modularization then regions near the extreme path points should map to regions close together. Regions far from the path's end points are free of this constraint.  That means the paths will be mapped into sort of "petal loops"  (Going forward we will call these loops the OM-petals) meeting at the central minimal cell. This fact will be important at the moment of interpreting the meaning of this spatial transformation. In Figure 1.d we can also see that under the transformation above the winding of the resulting loop around the minimal cell can be a Natural Number (n) larger or equal to 1. In order to differentiate these winding loops from OM-petals we will call them contractible OM-loops. In a 2-D visualization as Figure 1.d this winding around the minimum cell is partially dependent on the projection of the 3-D spacial paths to a 2-D view. However the maximum winding number is not dependent of this projection. In  at least one of the ways to perform this 2-D projection one gets a maximum possible such winding number. In any case the resulting general topology will be always the one of a Knot. Going forward we will call it the OM-Knot.  This OM-Knot is composed of one (or more) OM-petals an a integer number n of contractible OM-loops.\\ \\

 We can make at this point another important  observation which we will use in the rest of this paper. The length of the paths connecting the extreme points gets transformed into a integer multiple of the minimal cell space size (the Planck  Length)
 
\begin{equation} \label{Eq4}
N \zeta_0
\end{equation}

This is due to the fact that the loop can be homotopycally deformed to a loop curled directly over the minimal cell. Therefore we get the following OM correspondence for one  position coordinate

\begin{equation} \label{Eq5}
\tilde{x}=  \frac{x}{\zeta_0} = N 
\end{equation}

Let's note that this variable N and the number n of contractible loops are in general different natural numbers. n is partially dependent on the geometry of the system we are considering but it can also depend on symmetries or other physical constraints. In this work we will almost ignore the variable n and use it only for visualization. This variable hides in reality a very deep quantum baggage and will be the topic of a future work. In this paper we will focus mainly in the variable N which has a pure geometric origin. \\
The details of what the OM-Knot "petals" do far from the central minimal size cell is not influencing paths integrals of holomorphic (or branch-holomorphic) functions. Only the path integration around the central cell contributes to the end result. 
We see here the first hint that the Theory resulting from the applying the OM correspondence to QM will lead us to a Mathematical Theory, more precisely a Number Theoretical framework. We will use the number N loosely in this paper. We won't consider it to be strictly a Natural number. For us it  will be just a numeric parameter which arises due to the background transformation introduced in the OM correspondence. We can allow it to vary as a real parameter for the purpose of the rest of this paper. However, we will see later that most of the OM-QM counterparts of the quantum statements adopt the form of Number Theoretical statements and N will naturally be identifiable as a Natural.  \\ \\ 

From the discussion above we get that the Space-Time  transformation defined above allow us to give an operational form to Step 3 in our correspondence as the following way to transform a general path integral of an arbitrary phase-space function f(x,p)  

\begin{equation}\label{Eq6}
\int_{C[\alpha - \beta]} f(x,p) \rightarrow  \int_{ L[\alpha - \beta] } \tilde{f}(\tilde{x},\tilde{p})
\end{equation}

where $ C[\alpha - \beta] $ is a path (or a set of them) connecting states centered at Space-Time points $ \alpha $ and $ \beta $, $ L[\alpha - \beta] $ is a "petal-like" Loop (or a set of them) connecting at a minimal cell of the transformed space at points $ \alpha-\epsilon $ and $ \alpha+\epsilon $.  $ \tilde{f} , \tilde{x}$ and $ \tilde{p}$ are the OM-QM corresponding objects to the integrated function, position and momentum. Here x should be understood as a vector in 3 dimensions though below we will work over one of its components. We should keep in mind that for Knots the number of spatial dimensions is very relevant but we will see we will be able to work with each coordinate separately for the sake of visualization and for reducing complexity. 

$ \tilde{x} $ is given by  Eq.\ref{Eq4}. $ \tilde{p} $ can be found easily knowing $ \tilde{x} $ and remembering the expression of the quantum momentum operator 

\begin{equation}\label{Eq7}
p=i\hbar \dfrac{d \Psi(x)}{dx}
\end{equation}

with $ \Psi(x) $ the Wave Function of the quantum system.

Applying the OM correspondence we get 

\begin{equation}\label{Eq8}
\tilde{p}=i 2 \pi \dfrac{d \tilde{\Psi}(N) }{dN}
\end{equation}

where $ \tilde{\Psi}(N) $ is the OM-QM counterpart of the Wave Function. Later we will find a concrete expression for this $ \tilde{\Psi}(N) $ . By now we only need to know that it is a function of N through its dependency on $ \tilde{x} $. \\ \\

We need one more element to be able to make progress in finding what corresponds to QM in the OM-QM side. Although we could start with a non relativistic version of QM we will instead work in a Relativistic quantum framework from the start. To that end, we will need to find the OM counterpart of the speed of light ( $ \tilde{c} $ ) as this is something which also gets mapped to certain mathematical object under the OM-correspondence. We know how distances get mapped according to Eq.\ref{Eq4}. In order to know how a speed gets transformed we need to understand how time gets mapped in the OM-QM side. In the definition of the OM correspondence we mentioned that this is essentially an holographic theory where we are reducing degrees of freedom via the identification of certain regions of space-time. In every holographic transformation time gets mapped to an spacial radial coordinate tangent to the target spacial dimensions \cite{de2003introduction} \cite{gao1999ads}. In Figure 1.2 we visualize how this looks for the case of our modularized space-time.

\begin{figure}[h]
	\centering
	\includegraphics[width=0.4\linewidth]{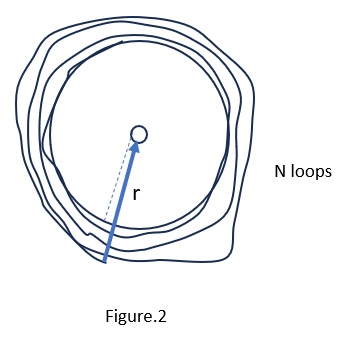}
\end{figure}

We can see time can be identified with the radius r in that picture and we should assume the transformed time flows in the direction of the central minimal cell. The length of the spacial dimensions is as we know $ N \zeta_0 $. Therefore we have 

\begin{equation}\label{Eq9}
|r|= \frac{N \zeta_0}{2 \pi}
\end{equation}

this can be assumed to be a minimum time similar as $ N \zeta_0 $ is a minimum length. Then we can write

\begin{equation}\label{Eq10}
|\tilde{c}| = \frac{N \zeta_0}{|r|}=2 \pi
\end{equation}

and finally to take into account the direction of this transformed radial coordinate to point inwards towards the central cell ($ \vec{r} = i |r| $), we finally find 

\begin{equation}\label{Eq11}
\tilde{c} = - 2i \pi
\end{equation}

\section{The micro and medium scale structure of OM-space}

In Figure 1 and in the previous section we got a visualization of the micro structure of the OM-space. Each loop of the fuzzy OM-knot leaves and enters the central minimal size cell in an almost parallel fashion. In general the loop can wind around the fuzzy center cell for a large number of full rounds. In a two dimensional visualization all those loops can be thought as a ring structure made of loops circling the central fuzzy center of the OM-Knot. However, we must notice that as we do not allow self crossing of the loop into itself the re-injection of the loop into the central region must occur in a direction above or below that ring structure. In the following figure we see a visualization of this medium scale structure.
\\ \\

\begin{figure}[h]
	\centering
	\includegraphics[width=0.7\linewidth]{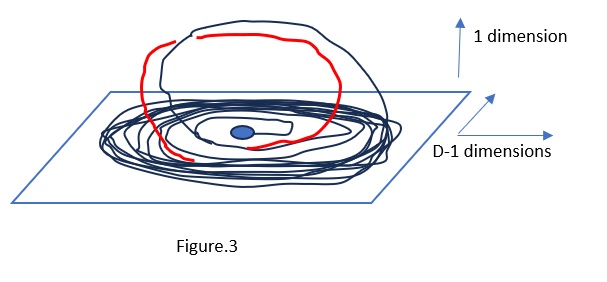}
\end{figure}

 It is very similar to an object known in the area of Dynamical Systems and Chaos as a Rosller attractor \cite{rossler1976equation} \cite{letellier2006inequivalent}. To be precise this is just and analogy we are proposing. The real target object has a very close resemblance of the Rosller attractor but its real structure can have some variations as we will see later. Nevertheless, going forward we are going to use the name Rosller-like geometry to describe the micro structure of the OM-pace. We propose that this  mathematical object is in effect a good representation of the medium scale structure of OM-space. It also shows a flow of loops around a central region in an almost two dimensional ring with re-injection flow along a perpendicular direction. In Chaos Theory the Rosller attractor araises in the phase space orbits of dynamical systems with dissipation due to interaction or coupling of two dimension in the system. The coupling between two of the system's dimensions is mapped into the re-injection of the flow of the orbits in one of those coupled dimensions into the general flow of orbits in the other dimensions. This causes a dimensional reduction in the resulting attractor of the orbits flow. The resulting object has a dimension between 2 and 3 with a fractal geometry. Locally this kind of strange attractors behave as a Baker Map. That is, each loop around central region stretches phase space volumes in one of the coupled dimensions while expanding them in the non coupled directions. The net effect if a volume preserving transformation for each cycle of orbits flow around the central region. A known characteristic of this kind of Baker flows is that the separation between distances ($ d_N $) between points in the flow increases in an universal exponential fashion involving the Lyapunov ($ \lambda $) coefficients of the loops flow 
 
 \begin{equation}\label{Eq12}
 d_{N+1} \propto d_{N+1} \exp{(\sqrt{\pi \lambda })}
 \end{equation}
 
 where for a Baker type map (locally a logistic like map), the Lyapunonv coeficient is equal to the Feigenbaund Constant ($ \delta = 4,66920160910299.... $ )
 
 The proportionality constant depends on the square of the Hausdorff dimension of the Rosller fractal. If the attractor is embedded in a three dimensional space then this proportionality constant will be the square of a number between 2 and 3 and rather close to 3 (the dimension of a locally logistic fractal embedded in 3D space). This will be important in another section later in this paper.
 
 It is worth noting that every observable of the system depending only on the scale we observe the system follows the same scaling universal behavior. For example, in many systems one of those properties depending only on the observation scale is Entropy. We will make use of this property later in the context of the OM correspondence.

\section{The OM-QM counterpart of the free particle Dirac Equation}
 
 Having found expressions both for $ \tilde{x} $, $ \tilde{p} $ and $ \tilde{c} $ we can now try to find the OM-QM counterpart of the Quantum Wave Function, Physical  Observables such as Energy and parameters such as the mass and spin. All of them get transformed to Number Theoretical objects under the OM correspondence. We work with all physical units in full display as in the OM correspondence we want to understand how all physical elements transform in the OM-QM side. \\ \\
 The free-particle Dirac equation has the following form 
 
 \begin{equation}\label{Eq13}
 (i \hbar \gamma^\mu \delta_\mu - mc) \Psi(x,t) = 0
 \end{equation}
 
 where the $ \gamma^\mu $ are the 4 by 4 Dirac Matrices, m is the particle's mass, c is the speed of light and $ \Psi(x,t) $ is a 4 components bi-spinor describing the particle quantum state. Plane waves solution to this equation make necessary the emergence of the particle spin. Step 3 of the OM correspondence in the case of the free-particle gives the simplest possible modulrization of the space-time as there is no topological or experimental constraint other than the ones implicit in the extremes of paths connecting the particle's initial position and end position. The fuzzy OM-Knot resulting from the identification of initial and end points produces a simple knot with only one petal with a minimum cell center. Let's apply our OM-correspondence to Eq.\ref{Eq13}. We obtain the following differential equation
 
 \begin{equation}\label{Eq14}
(i 2 \pi (\gamma^0 \dfrac{\delta}{\delta N_0} ) - i 2 \pi (\gamma^{1,2,3} \dfrac{\delta}{\delta N_{1,2,3}} ) ) \tilde{\Psi}(N_{0,1,2,3}) =  \tilde{m}\tilde{c} \tilde{\Psi } (N_{0,1,2,3}) 
\end{equation} 
 
 Here the numeric parameters $ N_{0,1,2,3} $ are the OM-QM side of each of the space time coordinates. We have also substituted the derivation in terms of the space-time coordinates by their OM-QM counterparts and $ \tilde{m} $ also transforms to a Number Theoretical expression as well as we saw the speed of light also does. We will see later that $ \tilde{m} $ is in fact a function of the numbers $ N_{0,1,2,3} $. The particle's state transforms to a 4-components vector where each of its components are functions of the numbers $ N_{0,1,2,3} $. \\
 Eq.\ref{Eq14} is the full OM-QM correspondent of the Dirac Equation. At this point we could try to find a plane wave solution and understand what is the  OM-QM version of  $ \tilde{\Psi } (N_{0,1,2,3} ) $ that is, of what is the OM-QM counterpart of the free particle state. The $\gamma$ matrices do mix the derivatives on the four $ N_i $ parameters. Therefore, finding a solution for one of the components of $ \tilde{\Psi } (N_{0,1,2,3} ) $ involve derivatives in all four $ N_i $ parameters. We will make here a simplification in order to find of a solution to Eq.\ref{Eq14}. Let's recall the topological meaning of the parameters $ N_i $ as winding of paths  in space-time around a minimum cell.  We can always find a topology of those "petal loops" where the projection of a 3-dimensional such knot according to any of the dimensions onto any of the 3 possible  2-dimensional views has the same winding around the central minimum cell (irrespective of the dimension we decide to project). That is, for the kind of knots we are dealing with the three "directions" $ N_i $ are essentially equivalent. We can then simplify Eq.\ref{Eq14} considering that all  four derivatives are equal. In a geometric context this would not be possible even when we consider all coordinates as equivalent. However, here we are dealing with pure mathematical objects, numbers in fact. They cannot be labeled as different if they are equal in value. For them we can then simplify Eq.\ref{Eq14} to this form for the 4 components of $ \tilde{\Psi } (N_{0,1,2,3} ) $ 
 
  \begin{equation}\label{Eq15}
 (i 2 \pi (\gamma^\mu ) \dfrac{\delta}{\delta N^{\mu}}  ) \tilde{\Psi}(N) =  - 2i \pi \tilde{m} \tilde{\Psi } (N) 
 \end{equation} 
 
 Substituting the Dirac $\gamma$ matrices Eq.\ref{Eq15} is equal to
 
   \begin{equation}\label{Eq16}
 \begin{pmatrix}
 i-1 \\
 1-i \\
 1-i \\
 i-1 
 \end{pmatrix} \dfrac{\delta}{\delta N}  \tilde{\Psi } (N) = \tilde{m} \tilde{\Psi } (N)
  \end{equation} 
 
By similarity with the Dirac equation solution, the numbers that appear in the r.h.s. before the derivative are the OM-QM counterparts of the spin. We see that the possible values of  $\tilde{s}$ are

\begin{equation}\label{Eq17}
\tilde{s} = \pm (i-1) 
\end{equation}

 The values of $\tilde{s}$ are independent of N as they come from the gamma matrices. Therefore the general solution to Eq.\ref{Eq15} is 
 
 \begin{equation}\label{Eq18}
 \tilde{\Psi } (N) =  e^{\frac{\int^{N_f}_{N_i} \tilde{m}(t) dt}{\tilde{s}}}
\end{equation}

 We can propose a reasonable guess for a solution of Eq.\ref{Eq15}. Let's start by assuming a solution of this form 

\begin{equation}\label{Eq19}
 \tilde{\Psi } (N) = e^{\phi(N)} 
\end{equation}

 If we assume an almost plane-wave solution type then $ \phi(N) $ has the meaning of a total phase difference after parallel transport over the fuzzy Knot we discussed in previous section. This is due to the fact that this phase difference corresponds to the comparison of the free particle states at the initial space-time point and at the final space-time point. As we saw before this phase comparison is done after parallel transport over the "petals" of the fuzzy OM-Knot generated by the OM correspondence. Comparison after parallel transport is equivalent to Curvature. In this case it is total curvature integrated over the whole OM-Knot "petals". If we call this curvature as R(N) then we have 
 
 \begin{equation}\label{Eq20}
 \tilde{\Psi } (N) = e^{\phi(N)} = e^{\int^{N_f}_{N_i} R(t) dt} = e^{\frac{\int^{N_f}_{N_i} \tilde{m}(t) dt}{\tilde{s}}}
 \end{equation}
 
 Essentially what we are saying is that a very reasonable guess  for the OM-QM  counterpart of the free particle quantum state is a topological invariant of the OM-Knot and concretely, the exponential of the total Knot curvature. This invariant is in general a function counting the statistics of possible crossing at the fuzzy center of the OM-Knot. From Eq.\ref{Eq20} we get 
 
 \begin{equation}\label{Eq21}
 \phi ' (N) =  R(N) =  \frac{\tilde{m}(N) }{\tilde{s}}
\end{equation} 
 
where $ \phi ' (N) = \dfrac{d \phi}{d N}$ 
  
We see here a first hint that the OM-QM statements gives very non trivial and interesting connections between elements coming from the QM side. In Eq.\ref{Eq21} we see an explicit relationship between the OM-QM counterparts of spin, mass and curvature. We cannot see that direct connection in the QM side but in the OM-QM side this arises explicitly.  \\ \\ 
Using the presence of the total OM-Knot curvature in the formula for $  \tilde{\Psi } (N) $ we can find a more explicit and meaningful form of that solution. From the work of Mazur it is known that the crossing statistics of fuzzy Knots are given by logarithms of the Prime Numbers \cite{mazur2012primes} \cite{morishita2009analogies} \cite{morishita2011knots}. To each fuzzy Knot one can assign a prime and to each prime one can assign a fuzzy Knot. Mazur gave the following correspondence 

 \begin{equation}\label{Eq22}
 \log (\textit{p}) =  vol(K)  
 \end{equation}

where \textit{p} is a prime number and vol(K) is the volume of the hyperbolic complement of a Knot \cite{callahan1998hyperbolic}. When talking about a fuzzy Knot that prime \textit{p} counts the crossing statistics at the fuzzy center. By the other side that volume is equal to the total curvature integrated over the Knot. Therefore, we have 

 \begin{equation}\label{Eq23}
\int^{N_f}_{N_i} R(t) dt = \log (\textit{p}) 
\end{equation}

At this point we make contact with Number Theory by recognizing that in the formula above we can use the von Mangoldt function $ \Lambda (\textit{p}) $ instead of $ \log (\textit{p}) $. The von Mangoldt function is defined by 

 \begin{equation}\label{Eq24}
\Lambda (\textit{p}) = \bigl\{ \begin{array}{cll}
\log(\textit{p})& if & N = \textit{p} ^k  \\ \\
0 & &otherwise 
\end{array}
\end{equation}

This is not a simple substitution as it takes into account the topological meaning of Eq.\ref{Eq20}. Substituting this into Eq.\ref{Eq20} we get the following possible solution for (each of the 4 components of) the OM-QM transformed Dirac Equation 

\begin{equation}\label{Eq25}
\tilde{\Psi } (N) = e^{(\sum_{q={\textit{p}^r}} \Lambda (\textit{N}) \delta (N-q))}
\end{equation}

As a side note, the exponent in this solution is equal to the derivative of the Chevyshev function ($ \psi_c $). The Chevyshev function is in turn equal to the logarithm of the Least Common Multiplier function of all numbers from 1 to N 

 \begin{equation}\label{Eq26}
\psi_c = \log(l.c.m.(1,...,N))
\end{equation}

Finally we find the following alternative form of $ \tilde{\Psi } (N) $

 \begin{equation}\label{Eq27}
\tilde{\Psi } (N) = exp[ \frac{l.c.m.'(1,...,N)}{l.c.m.(1,...,N)}]
\end{equation}

This gives an interesting interpretation to the solution of the OM-QM transformed Dirac Equation as a kind of focus or convolution formula for all scales from 1 to N. Let's remember that N in our framework represent scales in terms of the Planck Length. Therefore, $ \tilde{\Psi } (N) $ is in fact the exponential of the logarithmic derivative of a scale function which is able to resolve at the same time all the scales from 1 to N.

\section{The OM-QM counterpart of Mass and  Energy}

We will focus at this point in using Eq.\ref{Eq25} for finding the OM-QM counterparts of other Quantum Mechanical observables or parameters. By substituting Eq.\ref{Eq25} into Eq.\ref{Eq16} we get the following form for the OM-QM counterpart of the ]mass parameter in the Dirac equation

 \begin{equation}\label{Eq28}
\tilde{m}(N) = \tilde{s} \sum_{q={\textit{p}^r}} \Lambda (\textit{q}) \delta (N-q)
\end{equation}

Let's find the OM-QM version of the other equation satisfied by each one of the 4 components of the bi-spinor solution to the Dirac equation, that is the Klein-Gordon equation. 

 \begin{equation}\label{Eq29}
(\frac{1}{c^2}\dfrac{\delta^2}{\delta t^2} - \Delta^2+\frac{m^2 c^2}{\hbar^2}  ) \Psi(\vec{r}, t) =0
\end{equation}

We will consider the plane waves type solutions which allows us to substitute the time derivatives side by an energy eigenvalues term. If we apply the OM correspondence and use the transformed Dirac Equation to this equation we get the following OM-QM form of the Klein-Gordon Equation

 \begin{equation}\label{Eq30}
\frac{\tilde{E} ^2 (N)}{(4 \pi^2)^2} \tilde{\Psi}(N) = (\tilde{s}^2 \dfrac{d^2}{dN^2} + \tilde{m}^2(N) ) \tilde{\Psi}(N)
\end{equation}

where $ \tilde{E} $ is the OM-QM counterpart of Energy which as well mass becomes a function of the parameter N. To get to this equation we have made use of the same approximation we used in the case of the Dirac equation which allowed us to remove the dependency on the three spatial coordinates and work in one dimension. Using Eq.\ref{Eq20} we obtain this alternative version of the transformed Klein-Gordon equation 

 \begin{equation}\label{Eq31}
\frac{\tilde{E} ^2 }{(4 \pi^2)^2} \tilde{\Psi}(N) = (\tilde{s} \tilde{m}' + 2 \tilde{m}^2 ) \tilde{\Psi}(N)
\end{equation}

where we used the notation $ \tilde{m}' $ as the derivative of $ \tilde{m} $ respect to N. To find an explicit formula for $ \tilde{m}' $ we can use Eq.\ref{Eq28} and the method of derivation under the sum sign 

\begin{equation}\label{Eq32}
\tilde{m}' =  \tilde{s} \sum_{q={\textit{p}^r}} ( \dfrac{d(\Lambda (\textit{q}^b) \delta (N-q^b) )  }{db})|_{b=1} = \tilde{s} \sum_{q={\textit{p}^r}}  (\dfrac{d(\Lambda (\textit{q}))  }{dq} \log(q)q + \Lambda (\textit{q}) ) \delta (N-q)
\end{equation} 

Let's see how the derivative of the von-Mangoldt function is related to the Riemann $\zeta$-function. A known identity for the logarythmic derivative of the Riemann $\zeta$-function says that \cite{lebiedz2020holomorphic}

 \begin{equation}\label{Eq33}
\frac{\zeta'}{\zeta} = \sum_{q={\textit{p}^r}} \frac{\Lambda(q)}{q} \delta (N-q)
\end{equation} 

Otherwise we can write

 \begin{equation}\label{Eq34}
\frac{\delta \ln (\zeta)}{\delta N} = \sum_{q={\textit{p}^r}} \frac{\delta \Lambda(q)}{\delta q} \delta (N-q)
\end{equation} 

That means that $ \Lambda(q) $ is the Moebius inverse of  $ (\ln (\zeta (q)))' $ . Then $ \Lambda'(q) $ is the Moebius inverse of $ \zeta $. This implies a duality relationship between those two functions. From this we conclude that the peaks of the r.h.s. in Eq.\ref{Eq32} will occur on the zeroes of the $\zeta$-function. Then Eq.\ref{Eq32} can be considered an statistical average of those zeroes. We finally arrive to the following expression for $ \tilde{m}' $ 

 \begin{equation}\label{Eq35}
 \tilde{m}' = m + \tilde{s} \sum_{q={\textit{p}^r}}  ( \sigma_q \log(q)q  ) \delta (N-q)
 \end{equation} 
 
 where $ \sigma_q $ are the imaginary part of the zeroes of the Riemann $\zeta$-function. The statistical average of the real part of those zeroes (equal to 1/2 if the Riemann hypothesis is true) is zero. 

Substituting this into Eq.\ref{Eq31} we get the following formula for $\tilde{E}$

 \begin{equation}\label{Eq36}
\frac{\tilde{E} ^2 }{(4 \pi^2)^2} = \tilde{s}^2 \sum_{q={\textit{p}^r}}   \sigma_q \log(q)q  \delta (N-q)  +\tilde{m} + 2 \tilde{m}^2 
\end{equation}

That means that the  OM-QM correspondent of Energy for the free particle is a simple function of the imaginary part of the zeroes of the Riemann $\zeta$-function. It has been long theorized that the imaginary part of the zeroes of the Riemann $\zeta$-function are the eigenvalues of some non-hermitean Hamiltonian \cite{berry1999riemann} \cite{sierra2011h} \cite{connes2019scaling}. Many promissing  Hamiltonians have been tried but a proof of such conjecture still is out of reach. Here we have arrived to a similar relationship in the context of the OM correspondence. This shows the Number-Theory we arrive at with the OM correspondence is highly non trivial and can give us very valuable information about the Quantum side. We see that problems which seem to be hard or mysterious in a pure Quantum realm seems to be mapped to Mathematical relationships which are much simpler to treat. We see here a first hint that the OM-QM side of the correspondence seems to be a common playground for both Quantum Physics and Mathematics where the interplay of concepts from both sides can give raise to unexpected relationships (or conjectured as in the case above). 

\section{The values of $ \tilde{m} $ and $ \tilde{E} $ }
Let's explore the formulas for $ \tilde{m} $ and $ \tilde{E} $ to see which values they can possibly have. From Eq.\ref{Eq28} we see that only the following values are possible for $ \tilde{m} $

\begin{equation}\label{Eq37}
\tilde{m}(N) = \bigl\{ \begin{array}{ccl}
0 & if & N \neq p^k , p=prime \\ \\
\frac{\tilde{s}}{N} & if & N= p^k
\end{array}
\end{equation}

Thus, only for N which is a power of a prime number $ \tilde{m} $ is not null. The values it take are coherent with it the already used formula relating OM-Knot curvature to $ \tilde{m} $

 \begin{equation}\label{Eq38}
\tilde{R} (N)=\frac{\tilde{m}}{\tilde{s}}
 \end{equation}

as $ \tilde{R} $ is precisely of the form $ 1/N$  $ $. \\ \\
Inserting those values of $ \tilde{m} $ into Eq.\ref{Eq36} we get the following possible values for $ \tilde{E}^2 $

\begin{equation}\label{Eq39}
\tilde{E}^2(N) = \bigl\{ \begin{array}{lcl}
0 & if & N \neq p^k , p=prime \\ \\
\tilde{P}_0^2 (\tilde{s}^2 \sum_{q={\textit{p}^r}}   \sigma_q \log(q)q~\delta (N-q)  + \frac{\tilde{s}}{N} + 2 (\frac{\tilde{s}}{N})^2) & if & N= p^k
\end{array}
\end{equation}

We cam rewrite the values in the case $ N= p^k $ in terms of the OM-Knot curvature in this way

\begin{equation}\label{Eq40}
\tilde{E}^2(N) = \bigl\{ \begin{array}{lcl}
0 & if & N \neq p^k , p=prime \\
\tilde{P}_0^2 (\tilde{s}^2 \sum_{q={\textit{p}^r}}   \sigma_q \log(q)q ~\delta (N-q)  + \tilde{s} \tilde{R}(N) + 2 \tilde{s}^2 \tilde{R}^2(N)) & if & N= p^k
\end{array}
\end{equation}

We can reduce further this expression noting that the sum in it selects only one of the possible summing elements 

\begin{equation}\label{Eq41}
\tilde{E}^2(p^k) = \tilde{P}_0^2 (\tilde{s}^2 \sigma_p p \log(p)   \Delta_p  + \tilde{s} \tilde{R}(p^k) + 2 \tilde{s}^2 \tilde{R}^2(p^k))
\end{equation}

where $ \Delta_p $ is a distribution version of the Dirac delta centered at p.

We will return to this important expression later in this work.

\section{The value of the Fine Structure Constant }

As an application of our correspondence let's find find the OM-QM correspondent of the Fine Structure Constant. The value of this fundamental constant cannot be calculated from first principles in Quantum Mechanics even when is form in terms of other physical constants appears in the spin-orbit contribution to the energy eigenvalues in atomic Physics. Its is a non-dimensional number whose inverse is close to the rational $ \frac{1}{137} $. One of the ways to define formally this constant is through the following limit 

 \begin{equation}\label{Eq42}
\alpha^{-2} = \lim_{n \rightarrow \infty} \frac{{E^2}_{n-1}-{E^2}_{n-2}}{{E^2}_{n}-{E^2}_{n-1}}
\end{equation}

where $ E_n $ are the atomic energy eigenvalues. By similarity we can define the OM-QM correspondent of the Fine Structure Constant by the following number

 \begin{equation}\label{Eq43}
\tilde{\alpha}^{-2} = \lim_{i \rightarrow \infty} \frac{\tilde{{E^2}}_{i-1}-\tilde{{E^2}}_{i-2}}{\tilde{{E^2}}_{i}-\tilde{{E^2}}_{i-1}}
\end{equation}

where $ \tilde{E}_n $ are the OM-QM correspondent of Energy defined in Eq.\ref{Eq39}. Let's calculate explicitly this number $ \tilde{\alpha} $. Looking at Eq.$ \ref{Eq36} $ and Eq.$ \ref{Eq28} $ we notice that terms consecutive in N  related to mass are the same for $ N \rightarrow \infty $. That means they will cancel in $ \tilde{\alpha} $. We only need to consider differences of consecutive terms of the form 

\begin{equation}\label{Eq44}
\gamma (N) = \tilde{s}^2 \sum_{q={\textit{p}^r}}   \sigma_q \log(q) q   \delta (N-q)
\end{equation}

Thus, we get the following formula for $ \tilde{\alpha}^{-2} $ 

 \begin{equation}\label{Eq45}
\tilde{\alpha}^{-2} = \lim_{n \rightarrow \infty} \frac{\gamma_{n-1}-\gamma_{n-2}}{\gamma_{n}-\gamma_{n-1}}
\end{equation}

We know the difference $ \sigma_n-\sigma_{n-1} $ tends to zero in the infinite. Thus we get the following limit for $ \tilde{\alpha} $

\begin{equation}\label{Eq46}
\tilde{\alpha}^{-2} = \lim_{N \rightarrow \infty} K  \frac{\log(N-1) (N-1)-\log(N-2) (N-2)}{\log(N) N -\log(N-1) (N-1)}
\end{equation}

where K is the proportionality constant already mentioned in Section 4. Taking into account that N in our setup corresponds to a scale in terms of the Plank Length, all terms in $ \tilde{\alpha}^{-2} $ depend only on the scale and also have the form of an entropy term. We found in section 4  that such functions in the moduli space we are working on do scale in a very particular way (Eq.12). Specially entropy like functions do scale in that way. Therefore we get the following result

\begin{equation}\label{Eq47}
\tilde{\alpha}^{-2} = K \exp{(\sqrt{\pi \delta})}
\end{equation}

where $ \delta $ is the Feingenbaum constant. We get finally the following valued for $  \tilde{\alpha}^{-1} $

\begin{equation}\label{Eq48}
\tilde{\alpha}^{-1} = \sqrt{K} \sqrt{\exp{(\sqrt{\pi \delta})}}
\end{equation}

\begin{equation}\label{Eq49}
 \sqrt{\exp{(\sqrt{\pi \delta})}} \approx 46.061512666666666666666666666667 ....
\end{equation}

From our discussion in Section 4, we know the constant K is related to the square of the Hausdorff dimension of the local structure of OM-space. The dimension of a Rosller fractal embedded in 3D is a number between 2 and 3 and rather close to 3. A value of $ \sqrt{K}=3 $ gives a value of $ \tilde{\alpha}^{-1} = 138,184538......$ which is already close to the experimental value. However if we consider K to be the square of the Rosller fractal dimension (D=2.974283562752.....) we get the value $ \tilde{\alpha}^{-1} \approx 137 $. 

\begin{equation}\label{Eq50}
\tilde{\alpha}^{-1} = D \sqrt{\exp{(\sqrt{\pi \delta})}}
\end{equation}

In essence we have proven that assuming covariance of the form of $ \alpha^{-1} $ under our correspondence allows us to deduce its experimental  numeric value and its value does not change from the Quantum Mechanical value side. Then the Fine Structure Constant is an invariant in our correspondence 

\begin{equation}\label{Eq51}
\tilde{\alpha} = \alpha
\end{equation}

Let's notice that $ \tilde{\alpha}^{-1} $ is equal to the Lyapunov coefficient of the Rosller-like structure of OM-space times a coefficient dependent on its dimension. The Lyapunov coefficient generates in the Rosller internal flow the conservation of volume in the space complementary to the Rosller geometry. We propose that the conservation of this dual or complementary volume is the OM-QM correspondent to conservation of probability during  QM evolution. By the other hand, $ \alpha $ is a quantum Amplitude. More precisely, the probability of absorption or emission of a photon by a charged particle. Quantum Amplitudes are always volumes of certain geometrical objects embedded in certain spaces. Then, we can think of $ \tilde{\alpha}^{-1} $ as related to the volume of the hyperbolic complement of the OM-Knot. Then it makes sense that its value is almost exactly the coefficient of expansion-contraction in the OM-knot which secures conservation of that volume.

\section{The OM-QM correspondent of the electron's charge and Electric Field }

As all physical quantities in the QM side of our correspondence have a OM-QM counterpart it is natural to try to find the OM-QM version of the electron's electric charge. This is related to $ \alpha $ via the formula

\begin{equation}\label{Eq52}
\alpha = \frac{e^2}{\hbar 4 \pi \epsilon}
\end{equation}

where e is the electron charge and $ \epsilon $ is the vacumm electric permeability. This expresion of $ \alpha $ comes from atomic physics and it contains the constant $ \epsilon $ (the vacuum electric permitivity) which cannot be determined from first principles and comes from experimental measurement. If we invert this expression and apply the OM correspondence we find 

\begin{equation}\label{Eq53}
\tilde{\alpha}^{-1} = \frac{ 8 \pi^2 \tilde{\zeta_0} \tilde{\epsilon}}{\tilde{e}^2}
\end{equation}

where $ \tilde{e} $,  $ \epsilon $ and $ \tilde{\zeta_0} $ are the OM-QM version of e, $ \epsilon $ and the Planck Length. As The Planck Length is 1 in the scales transformation implied in our correspondence we can assume $ \tilde{\zeta_0} = 1 $

Then from Eq.\ref{Eq50} we get

\begin{equation}\label{Eq54}
\tilde{\alpha}^{-1} = \frac{ 8 \pi^2 \tilde{\zeta_0} \tilde{\epsilon}}{\tilde{e}^2} = D \sqrt{\exp{(\sqrt{\pi \delta})}}
\end{equation}

which after rearrangement give us the following formula for the quotient

\begin{equation}\label{Eq55}
\frac{\tilde{e}^2}{\tilde{\epsilon}} = \frac{8 \pi^2}{D \sqrt{ \exp{(\sqrt{\pi \delta})} }}
\end{equation}

we will disentangle this quotient in the l.h.s. making an educated guess of which part of the r.h.s. corresponds to denominator and numerator. We arrive to the following OM-QM versions for e and $ \epsilon $

\begin{equation}\label{Eq56}
\tilde{e} = 2 \pi
\end{equation}

\begin{equation}\label{Eq57}
\tilde{\epsilon} = \frac{D}{2} \sqrt{ \exp{(\sqrt{\pi \delta})} } = \frac{\tilde{\alpha}^{-1}}{2}
\end{equation}

The value of $ \tilde{e} $ can be interpreted as the residue of a potential of the form $ 1/z $ integrated around a loop around the minimum cell. What is more surprising is that the OM-QM counterpart of $ \epsilon $ turns out to be half the OM-QM counterpart of $ \alpha^{-1} $. As we saw previously $ \alpha^{-1} $ is related to the Lyapunov coefficient of the Rosller-like microstucture of OM-space. Then $ \tilde{\epsilon} $ also is. This seems to indicate a totally  unexpected connection between the OM-QM counterpart of $ \tilde{\epsilon} $ and the conservation of volume in the OM-speace Rosller-like geometry. Following we will see two arguments why this connection makes sense. First we can think in the following physical argument. In the previous section we showed that $ \tilde{\alpha}^{-1} $ is also related to the same dual volume conservation. Also, Physically, $ \tilde{\alpha}^{-1} $ and  $ \tilde{\epsilon} $ are talking about the same property, the amplitude of the coupling of photons to particles and to the vacuum respectively. We see that their OM-QM correspondents fall into the exact same numerical value. \\ \\ \textbf{We conjecture that this is an example of a general behavior of the OM-correspondence. Similar properties in the QM realm which differ only on whether they are talking about particles or about vacuum space will be mapped into the same mathematical object in the OM-QM side. We propose this is fact a central principle in the OM-correspondence, an analog to the Equivalence Principle in General Relativity. } \\ \\ Now let's see a geometric argument on why this connection is plausible. Let's analyze the behavior of the (classical) Electric Field under the OM  correspondence. In the following figure we show how the Electric Field of the electron changes after the  OM correspondence

\begin{figure}
	\centering
	\includegraphics[width=0.7\linewidth]{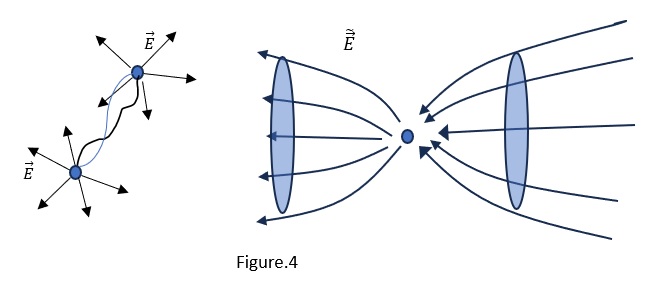}
\end{figure}

We start with a charged particle in normal space-time with classical Electric Field pointing outwards the particle's position both in the initial and final positions in the evolution path. When we modularize this space all the space-time regions near the extreme of the path are mapped to the region near the minimal cell and then in the OM-space all the transformed Electric Field lines will come out radially from the region near the minimal cell. As our modularization implies all lines have to enter this region in almost parallel fashion (and pointing in the same direction!) that means that one set of lines will be inwards to this region an another will be outwards to it. Moreover, due to the gluing of this central region we will have a mirror symmetry between the transformed fields entering and leaving this region. Thus, the transformed Electric Field is not spherical but rather hyperbolic in symmetry as the point in the infinite also has to be identified in the modularization. The transformed Electric Field looks like a Magnetic Field in an Hyperbolic Volume around the central region of the OM-Knot. If we identify this hyperbolic region where the Electric Fields lands with the hyperbolic complement of the OM-Knot itself then we have a very satisfying visualization of why the the OM-QM versions of the vacuum electric permeability is also so intimately and exactly related to the flow of volumes in the Rosller-like OM-space.

It is worth mentioning that the emergence of a sort of the Electric-Magnetic duality described above is a common feature of Field Theories formulated on moduli spaces \cite{verlinde1995global}. Thus it is not strange seeing this also appearing in our correspondence. What is new here is that we can identify the target "Magnetic lines" as occupying a very meaningful volume in the moduli space: the hyperbolic complement of the OM-Knot.

\section{The OM-QM correspondent of Heisenberg's Uncertainity relation }

Let's recall the form of the only non null possible values of $ \tilde{E}(N) $

 \begin{equation}\label{Eq58}
 \tilde{E}^2(p^k) = \tilde{P}_0^2 (\tilde{s}^2 \sigma_p p \log(p)  ) \Delta_p  + \tilde{s} \tilde{R}(p^k) + 2 \tilde{s}^2 \tilde{R}^2(p^k))
 \end{equation}

where $ \Delta_p $ is a distribution version of the Dirac delta centered at p. 
Let's analyze this expression in two limits. First let's see it in the limit $ p \rightarrow \infty $

\begin{equation}\label{Eq59}
\lim_{p \rightarrow \infty} \tilde{E}^2(p^k) \rightarrow \tilde{P}_0^2 (\tilde{s}^2 \sigma_p p \log(p)  ) \Delta_p 
\end{equation}

We can notice that the expression $ p \log(p)  ) \Delta_p $ is dependent only of the scale p and we know how those quantities transform under change of p (Eq.\ref{Eq12}). Then if we calculate the difference of $ \lim_{p \rightarrow \infty} \tilde{E}^2(p^k) $ between two consecutive values of p we get the following formula 

\begin{equation}\label{Eq60}
 (\Delta (\tilde{E}(p^k)))^2 = \tilde{P}_0^2 \tilde{s}^2 D^2 \exp{(\sqrt{\pi \delta})} (\sigma_{p+1} - \sigma_{p})
\end{equation}

Recalling the relationship between $ D^2 \exp{(\sqrt{\pi \delta})} $ and $\tilde{\alpha}^{-2}$ we get the following in this limit 

\begin{equation}\label{Eq61}
\Delta (\tilde{E}(p^k)) (\frac{\tilde{\alpha}}{(\Delta(\sigma))^{1/2}}) = \tilde{P}_0 \tilde{s}
\end{equation}

We argue that this is the OM-QM correspondent to the Heisenberg Uncertainty relation for Energy and Time. In the r.h.s. of this equation we have the correspondent of the Plank constant. and in the l.h.s. a product of variations. One of them is the variation of the  OM-QM correspondent of Energy. As we are assuming covariance under our correspondence then the quantity $ \frac{\tilde{\alpha}}{(\Delta(\sigma))^{1/2}} $ must be the OM-QM correspondent of variations of certain times. Thus, $ (\Delta(\sigma))^{1/2} $ must be the OM-QM correspondent of variations of certain frequencies. This give us as a bonus the interpretation of the imaginary parts of the zeroes of the Riemann $\zeta$-function as the OM-QM correspondent of frequencies associated to $\tilde{E}$.

\section{The OM correspondent Einstein's Gravitational field equation }

Equation Eq.\ref{Eq36} is the OM-QM correspondent to Klein-Gordon's equation for a free particle, a second order relationship between Energy, momentum and mass. In the OM-QM side this becomes a relationship between the OM-QM correspondent of Energy, one term related to the OM-Knot curvature and an additional term which we will discuss in more detail below. In this section we will work with the values of this equation for finite values of the prime p. It is known that getting a first order relationship between Energy observable, momentum and mass was the starting point for Dirac for finding the quantum state Equation for the free particle in a relativistic context. The prize to pay for getting a first order relationship was that the coefficients in the equation no longer could be scalars but needed to be matrices, the Dirac matrices. We will see that in the context of our correspondence we also can find a first order relationship starting with Eq.\ref{Eq41}. \\ \\
Let's start by taking the difference of $ \tilde{E}^2 (p) $ between two consecutive but finite primes p. We get the following relation

 \begin{equation}\label{Eq62}
\Delta (\tilde{E}) = \tilde{P}_0 (\tilde{s} \Delta (\tilde{R}) + 2 \tilde{s}^2 \Delta (\tilde{R}^2) + \tilde{s}^2 \Delta (\sigma_p p \log(p)   \Delta_p)) ^{ 1/2}
\end{equation}

where we have approximated $ \Delta (\tilde{E} ^ 2) $ by $ [\Delta (\tilde{E})]^2 $

The quadratic term $ \Delta (\tilde{R}^2) $ dominates over the linear one $ \Delta (\tilde{R}) $ for finite p. It can also be approximated by $ [\Delta (\tilde{R})]^2 $. We can then write 

 \begin{equation}\label{Eq63}
\Delta (\tilde{E}) = \tilde{P}_0 \tilde{s} (2 [\Delta (\tilde{R})]^2 + \Delta (\sigma_p p \log(p)   \Delta_p)) ^{ 1/2}
\end{equation}

The only way we can reduce this equation to a first order equation is if the factor under the square root is a square, that is whether there exists a Pythagorean relationship between $ 2 [\Delta (\tilde{R})]^2 $ and $ \Delta (\sigma_p p \log(p)   \Delta_p) $. From Number Theory we know that this question is related to the possibility to have unique complex prime factorization in a Number Field made up of elements of the form 

 \begin{equation}\label{Eq64}
\sqrt{2} \Delta (\tilde{R}) + i \sqrt{\Delta (\sigma_p p \log(p)   \Delta_p)}
\end{equation}

And that is tied to the existense of a number close to an integer Number and of the form $ \gamma^2 \exp{(\sqrt{\pi \delta})} $ whith $ \gamma $ and $ \delta $  positive numbers. The two squares in the Pytagorean relationship should sum up to precisely this number. In the case of the elements inside the square root in Eq.\ref{Eq64} we know that both elements depend only of the scale and under changes in the number p they change according to the Logistic relationship $ D^2 \exp{(\sqrt{\pi \delta})} $ . And we saw in previous sections that this number is very close to the square of the inverse of the Fine Structure Constant. It s also the factor by which volumes in the OM-space transforms under changes in p. This  proves that we can write Eq.\ref{Eq63} as the product of two mutually conjugate complex equations of the form

 \begin{equation}\label{Eq65}
\Delta_+ (\tilde{E}) = \tilde{P}_0 \tilde{s} (\sqrt{2} \Delta (\tilde{R}) + i  \sqrt{ \Delta (\sigma_p p \log(p)   \Delta_p)}  )
\end{equation}

We want to emphasize how remarkable this relationship is. Its existence is dictated Mathematically by the way OM-space transforms under changes in p which in turn determines the value of the inverse of the Fine Structure Constant. It is precisely that value (so close to an integer) what makes this relationship a valid one. It is a first order relationship between $ \tilde{E} $ ,  $ \Delta (\tilde{R}) $  and another term. Similar to the case in the original Dirac's result we have here a price to pay for the passing to a first order relation. In the Dirac equation there was the appearance of Dirac gamma matrices. In our case it is the occurrence of an imaginary unit in front of that last term $ \sqrt{\Delta (\sigma_p p \log(p)   \Delta_p)} $ . \\ \\
At this point we make connection with Einstein's Field Equation in General Relativity as they are also a relationship between Energy and Curvature. They also include an extra term historically called the Cosmological Constant term, a term which modernly we associate to the action of Dark Energy. We are proposing that Eq.\ref{Eq65} is indeed the OM-QM correspondent to Einstein Field Equation for the Gravitation Field in their following form

 \begin{equation}\label{Eq66}
T = \frac{c^4}{8 \pi G} ( (1-\frac{D}{2}) R + D \Lambda )
\end{equation}

where T is the net value of the Mass-Energy tensor, G is the Gravitational Constant, D is the space Dimension and $ \Lambda $ is the Cosmological Constant term. \\ \\
Applying the OM correspondence to this equation we get 

 \begin{equation}\label{Eq67}
\tilde{T} = \frac{\tilde{c}^4}{8 \pi \tilde{G}} ( (1-\frac{D}{2}) \tilde{R} + D \tilde{\Lambda} )
\end{equation}

substituting $ \tilde{c}^4 = 8 \pi^2 $ we arrive to the following formula 

 \begin{equation}\label{Eq68}
\tilde{T} = \frac{2 \pi^3 }{ \tilde{G}} ( (1-\frac{D}{2}) \tilde{R} + D \tilde{\Lambda} )
\end{equation}

Comparing Eq.\ref{Eq68} and Eq.\ref{Eq65} we get the following relationship 

 \begin{equation}\label{Eq69}
\frac{2 \pi^3 }{ \tilde{G}}  (1-\frac{D}{2})  = \sqrt{2} \tilde{s} \tilde{P}_0
\end{equation}

from which we get the following expression for the OM-QM correspondent to the Gravitational Constant

 \begin{equation}\label{Eq70}
\tilde{G} = \frac{\pi (1-\frac{D}{2})}{2^{2/3} \tilde{s}}
\end{equation}

We also get the following from the comparison 

 \begin{equation}\label{Eq71}
\frac{2 \pi^3 }{\tilde{G}} D \tilde{\Lambda} = i \tilde{s} \tilde{P_0} \sqrt{\Delta (\sigma_p p \log(p)   \Delta_p)} 
\end{equation}

from which we get the following form of the OM-QM correspondent to the Cosmological Constant

 \begin{equation}\label{Eq72}
\tilde{\Lambda} = \frac{i \pi }{\sqrt{2}} \frac{(1-\frac{D}{2})}{D}\sqrt{\Delta (\sigma_p p \log(p)   \Delta_p) }
\end{equation}

\section{The OM-QM correspondent of Holography  }
Holography is one of the most exiting discoveries in Physics in the recent decades. Starting with the original idea from Gerard T'Hooft and followed by the insights of Susskind and Maldacena it has already got experimental applications in several areas of Physics \cite{susskind1995world} \cite{bousso2002holographic}. In essence Holography states that every Classical Physical Theory with Gravity in a certain D+1 volume (in a certain space) can be mapped to Quantum Theory without Gravity on the D dimensional boundary of that same D+1 volume. Quantum Mechanics is an holographic Theory, so is General Relativity. That is why it is generally believed that Holography will be an essential component of any future Quantum Gravity Theory. One explicit toy model of such Theory is the so called AdS/CFT correspondence by Maldacena \cite{hawking2001desitter}. Here AdS is a classical gravity theory living in a finite hyperbolic space (AdS) and CFT is the holographic quantum dual theory living in the boundary of AdS. This correspondence gives a very convenient toy-model for visualization of the details of the holographic mapping to the boundary and a how the quantities inside the gravitational bulk do get mapped to quantum observables in the boundary. Gravity for example must become null in the boundary. \\ \\More recently Maldacena and Susskind have proposed a physical interpretation of entanglement as the holographic dual of an Einstein-Rossen bridge, the so called ER=EPR Theory \cite{susskind2016copenhagen}. Entanglement being the the source of so many of the apparently counter intuitive properties of QM would then be the shadow of an object in a higher dimensional gravitational dual. \\ \\
Let's see how holography gets mapped under the OM correspondence. We start by analyzing the formula for $ \tilde{G} $

\begin{equation}\label{Eq73}
\tilde{G} = \frac{\pi (1-\frac{D}{2})}{2^{2/3}  \tilde{s} }
\end{equation}

Let's remember that D is the dimension of the Rossler-like micro-structure of the OM-space. It's a fractal dimension close to 2.9. Let's see how is the behavior of $ \tilde{G} $ when we vary this dimension D towards 2. It indeed  becomes null as it should if we were dealing with a proper holograhic limit. Let's now analyze the denominator. Here we have $ \tilde{s} $, which for us is just a complex number by now arising in the our transformed Dirac equation. Using that equation we can rewrite the following 

\begin{equation}\label{Eq74}
\tilde{G} = \frac{\pi (1-\frac{D}{2}) \tilde{R}}{2^{2/3}  \tilde{m} }
\end{equation}

where we have substituted $ \tilde{s} $ by the 	quotient of the OM-Knot curvature and $ \tilde{m} $. We can interpret then $ \tilde{G} $ as how much D-1 dimensional curvature is generated by $ \tilde{m} $. Here we are off course using the fact that we have imposed covariance of the meaning of the mathematical quantities we have in our equations respect to the physical quantities in the quantum side of our correspondence. That is, the purely mathematical equations in the OM-QM side can shed some meaningful information about the observables in the $ P_0 $ side and vice versa.   \\ \\
Rearranging Eq.\ref{Eq73} we get 

\begin{equation}\label{Eq75}
\tilde{s} \tilde{G} = \frac{\pi (1-\frac{D}{2})}{2^{2/3} }
\end{equation}

in the r.h.s. of this relation we have a pure number. That tell us that $ \tilde{m} $ has a duality relation with $ \tilde{G} $. $ \tilde{m} $ must represent something dual to a curvature in D-1 dimensions. Eq.\ref{Eq75} is an holographic relation too. Therefore we propose this give us an interpretation of $ \tilde{s} $ as a curvature in higher dimensional OM-space. If we believe in the covariance imposed into our correspondence then this give us the following interpretation of the spin observable in QM  \\  \\
\textbf{The quantum mechanical observable we call spin in the usual QM is the D-1 shadow of the curvature of the OM-Knot in a dimension perpendicular to all those D-1 dimensions we can observe. The spin in the OM correspondence can be interpreted as a curvature in a higher dimension we cannot observe directly. We just see its shadow as quantum mechanical effects. This dimension is coupled to the dimensions we can observe}\\ \\ 
Going forward we will adopt the following notation. We will call $ \tilde{R}_H $ this curvature in this higher dimension we cannot observe directly and we will use it instead of $ \tilde{s} $ in our equations. We will call $ \tilde{R}_L $ the curvature in the D-1 dimensions which we can observe directly. Thence, we can rewrite the transformed Dirac equation in this suggestive way 

\begin{equation}\label{Eq76}
\tilde{R}_H \tilde{R}_L = \tilde{m}
\end{equation}

Let's visualize this in the Rosller-like micro structure of the OM-space. In the following figure we can see the parts of the OM-Knot in the D-1 dimensions we can observe and the "reinsertion flow" of the knotted paths which must necessarely go outside of the D-1 volume in order to achieve the Rosller reinsertion.

\begin{center}
	\centering
	\includegraphics[width=0.8\linewidth]{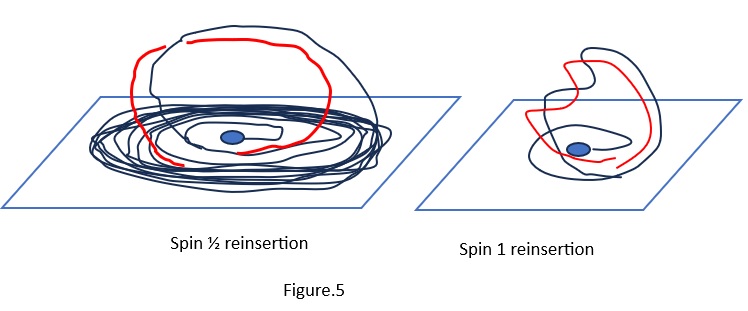}
\end{center}

The curvature of this "reinsertion flow" is the spin. This also explains the quantized values of spin we observe in QM. The reinsertion flow can flip only a integer multiple of 1/2 times in its trajectory in the high dimension. That is why we see a total shadow effect in the lower dimensional in the form of the appearence of a quantity which can only have values which are multiples of 1/2. \\ \\

As we see, if we work in a QM setup we are really not seeing the full picture but only the holographic projection of the D dimensional Rosller OM-space into D-1 dimensions. As we saw in the case of the spin, quantization in D-1 dimensions is an effect of the coupling of that extra dimension to the D-1 ones we can observe directly. The spin and other quantized quantities can be then viewed as a window to that higher dimension we cannot observe. \\
Let's then try to generate a framework where we see an unified picture in all the D dimensions of the Rosller OM-space.  \\ \\
We start by noting that the total curvature in the full OM-space can be written in this way

\begin{equation}\label{Eq77}
\tilde{R} = \tilde{R}_H  + \tilde{R}_L
\end{equation}

Let's take the square of this total curvature

\begin{equation}\label{Eq78}
\tilde{R}^2 = \tilde{R}_H^2 + \tilde{R}_L^2 + 2 \tilde{R}_H \tilde{R}_L
\end{equation}

and using Eq.\ref{Eq76} we get

\begin{equation}\label{Eq79}
\tilde{R}^2 = \tilde{R}_H^2 + \tilde{R}_L^2 + 2 \tilde{m}
\end{equation}

We saw before that the possible values of R are of the form $ \frac{1}{N} $. Then $ R^2 = 1/N^2 $. As R is curvature in the hyperbolic complement of the OM-Knot then the inverse of the square of the curvature radius is equal the variation of D-1 dimensional Area in that hyperbolic complement. Visualization of this fact is difficult in OM-space. However an analogy with the AdS/CFT visualization of the relations between Area in AdS space and length in the boundary of that space is a good analogy for any holographic theory. We can then write

\begin{equation}\label{Eq80}
\tilde{R}_L^2 = \Delta\tilde{A}_L
\end{equation}

where $ \Delta \tilde{A}_L $ is a variation of Area in the hyperbolic complement of the OM-Knot in the D-1 projection. Using this we can rewrite Eq.\ref{Eq79} in this form

\begin{equation}\label{Eq81}
\Delta\tilde{A}_L = \tilde{R}_H^2 - \tilde{R}^2 + 2 \tilde{m}
\end{equation}

The difference $ \tilde{R}_H^2 - \tilde{R}^2 $ is equal to minus the difference of OM-Area after one single flow of the OM-Knot in the higher dimension. We will note this in this by $ \Delta\tilde{A}_H $. Thus we arrive to this result 

\begin{equation}\label{Eq82}
\Delta\tilde{A}_L = - \Delta\tilde{A}_H + 2 \tilde{m}
\end{equation}

\textbf{In itself this is a remarkable relationship between OM-mass and variation of the total OM-Area. It provides an holographic interpretation of $ \tilde{m} $ as a variation of the OM-QM correspondent of Area of the hyperbolic complement of the OM-Knot.} \\ \\ 

We can give a more explicitly holographic form of this formula recalling that in any holographic theory there is a universal relationship between the Area in the D dimensional Theory and the entanglement entropy in the quantum theory living in the boundary D-1 space. This is the so called Area Law of holography. This relationship first appeared in Black Hole physics in the works of Jacobson and Hawkings. In its pure Quantum Mechanical form we can write it like this

\begin{equation}\label{Eq83}
\Delta A_L = - \Delta S_L 4 G
\end{equation}

Where $ S_L $ is the entanglement entropy in the boundary of the D dimensional space and G is the Gravitation constant. Its OM-QM form can be written as

\begin{equation}\label{Eq84}
\Delta \tilde{A}_L = - \Delta \tilde{S}_L 4 \tilde{G}
\end{equation}

Substituting it in Eq.\ref{Eq82} we get the following relationship 

\begin{equation}\label{Eq85}
\Delta \tilde{S}_L =  \frac{\Delta\tilde{A}_H}{4 \tilde{G}} -  \frac{\tilde{m}}{2 \tilde{G}}
\end{equation}

This, we propose, is the OM-QM version of another universal holographic relation, the Ryu-Takayanagi relation which (in its pure Quantum Mechanical version) it states the following relationship between Area, space related Entanglement Entropy and matter related entropy 

\begin{equation}\label{Eq86}
\Delta S_L =  \frac{\Delta A_H}{4 G} -  (matter-entropy-term)
\end{equation}

The emergence of such a relationship strongly supports the idea that OM-QM is also a proper holographic Theory and holography is covariantly preserved in the OM-correspondence. \\ \\

 \section{Conclusions and outlook}
 In this work we have proven that we can construct a very special spatial background by identifying regions of normal space where observation (or measurement) are implied (OM-space). We have shown that Quantum Mechanics over this background reduces to a highly non trivial and interesting Number Theory (OM-QM) if we simultaneously make use of a reparametrization replacing the Planck constant by the quantity $ 4 \pi^2 \zeta_0 $ (where $ zeta_0 $ is the Plank Length) or equivalently replacing the Plank Momentum ($ P_0 $) by the number $ 4 \pi^2 $. Although we are transforming QM statements into pure Mathematical ones we are demanding this transformation to preserve the relationships among the original physical objects into the corresponding relationships among the target mathematical objects. This requirement of covariance causes that all objects in the QM side do transform under this correspondence. Elements like mass, energy, wave function, electric charge, spin, position, time, speed of light and all their derived objects become Number Theoretical objects in the OM-QM side. The functions and values they adopt as well as the transformed QM equations give a lot of new and non trivial insight about the objects the Physical side. We found the OM-QM equivalent to the Dirac and Klein-Gordon equations. From them we have learned that the OM-QM correspondent to Energy of a free particle are simple functions of the zeroes of the Riemann $ \zeta $-function. Similar relation between those zeroes and quantum energy eigenvalues been suggested in the past by several authors and here it appears naturally.\\
 The exploration of the meso and micro structure of OM-space lead us to a toy model local representation via a geometrical object similar to a Rolller attractor which exposes the dimensional reduction implicit in the modularization of OM-space. One of the main conclusions of this work is that topics which seems to be hard or of unknown origins in the QM side become easy to treat or to interpret in the OM-QM side. The mathematical insight we can get in the OM-QM side can shed some light on the nature or relationship among physical objects in the QM side (and vice versa!). This Rosller like structure allowed us to propose a first principles deduction to the exact numerical value of the Fine Structure Constant. Based on this toy model we were also able to find the OM-QM analogues of the Heisenberg relationship for Energy and Time an the Einstein Gravitation Field Equations emerging as two limits from the OM-QM analog to the Klein Gordon equation.  Although the intention in this work has been to stick to basic Quantum Mechanics not mixing when possible into the hard topics o Quantum Gravity we think that OM-QM is in fact some sort of Quantum Gravity Theory. It is so in the same sense that for example Maldacena's AdS/CFT is a full Quantum Gravity toy model. In our case we were not attempting to insert Gravity in the discussion at all but it appeared naturally into OM-QM through the implications of the OM-space Rosller micro structure into equations relating Energy and curvature of mathematical objects in the OM-QM side.  It is exactly the internal geometrical flow inside this Rosller like structure which allows us to pass from a second order Klein-Gordon OM-QM equation to a first order Einstein Gravitation field OM-QM equation. This is a highly non-trivial result and unique to the a Rosller like structure of OM-space. No other kind of geometrical structure would allow for such a result. The fact that holography is preserved and manifest in the OM-correspondence is also a sign that OM-QM is a type of Quantum Gravity toy model. It not only proves good internal  consistency but it also provides new insights like an interpretation of spin of a free particle as higher dimensional curvature of the Rosller OM-space micro structure. Also the OM-QM mass correspondent gets an interpretation as variations of the OM-Area of a geometrical object in the OM-QM side. We demonstrated that this in disguise the OM-QM version of the Ryu-Takayanagi universal relation for holographic theories. Another way to see OM-QM as a Quantum Gravity Theory is that based on analysis of the OM correspondents of similar QM quantities we were able to conjecture an OM-version of the Equivalence Principle in General Relativity. In essence we found that statements or quantities in the QM side which differ only in that one is speaking about particles and the other about vacuum space are mapped in exactly the same mathematical object in the OM-QM side. This important conjecture, its application and meaning is an obvious future venue of investigation for the OM-correspondence. We think that it could play a similar central role in the OM-correspondence as the Equivalence Principle plays in General Relativity. \\ \\
 Our intention in the paper was introducing the OM-correspondence with an absolute minimum of assumptions or without entering in side discussions on topics which are hard to grasp in physical or mathematical terms. Nevertheless, we see that the richness and cross connected nature of OM-QM is so large that one can speculate with the potential applications of it to some of those harder problems. As we mentioned above, exploring the value of OM-QM as a valuable toy mode for Quantum Gravity would be one of the potential future directions. In addition in as a work following up to this one we will apply the OM-correspondence to the following topics which in standard QM are hard or have some mysterious aspects to them :
\begin{itemize}
 \item  The State reduction principle (or Wave function collapse)
 \item The measurement problem in QM
 \item The EPR experiment interpretation 
\end{itemize}
 As potential future work venues we will explore the possibility of an experimentally observable consequence of OM-QM in the Cosmic Microwave Background Map or in the observed  matter clustering in the Early Universe. Applying the OM-correspondence to Black Hole physics could be also a potentially great area for exploration. \\
 Finally, we want to bring attention on what is probably the most striking and new aspect of the OM-correspondence and that is the reparametrization we have called the $P_0 - 4 \pi^2$ correspondence. We have explicitly avoided entering in any deep interpretation on why this works so well together with the OM-space structure. This reparametrization seems to balance the effect in QM of the OM-space modularization in such a way that by requiring covariance in the form of equations we get in the mathematical side objects which have very desirable mathematical relationships if they would be correspondents to their physical counterparts. This is most evident in the case of a pure numeric quantity which is also crucial in QM, the Fine Structure Constant. The OM-correspondence maps it into something which is also very central to OM-QM. Also the relationship between the OM-QM values of the free particle Energy and the zeroes of the Riemann $ \zeta $-function is possible only through this parametrization. We suspect the $P_0 - 4 \pi^2$ correspondence says something deep about the relationship between QM and Mathematics. OM-QM itself seems to be a sort of common playground where both QM and parts of Mahematics can be mapped one to each other. The search for this kind of relationships between QM and several areas of Mathematics has been a very active area of research in last decades. The Langlands Program is one of those ideas in Mathematics trying to build bridges with QM at a fundamental level. We think that the OM-correspondence may be studied from a more formal mathematical point of view which could shed some light on the deeper meaning of the $P_0 - 4 \pi^2$ correspondence. We propose that studying the potential relevance of the OM-correspondence for the Langlands Program in Mathematics could be a high value potential future research venue.

 \bibliographystyle{unsrt}
 \bibliography{bibliogra1}

\end{document}